# Dust particle charge in plasma with ion flow and electron depletion

Angela Douglass, Victor Land, Lorin Matthews, and Truell Hyde[a]

*Center for Astrophysics, Space Physics, and Engineering Research,*

*Baylor University, Waco, TX, USA 76798-7310, www.baylor.edu/CASPER*

The charge of micrometer-sized dust particles suspended in plasma above the powered electrode of radio-frequency (RF) discharges is studied. Using a self-consistent fluid model, the plasma profiles above the electrode are calculated and the electron depletion towards the electrode, as well as the increasing flow speed of ions towards the electrode, are considered in the calculation of the dust particle floating potential. The results are compared with those reported in literature and the importance of the spatial dust charge variation is investigated.

[a]Electronic mail: Truell_Hyde@baylor.edu

Dust particle charge in plasma with ion flow and electron depletion

**I. INTRODUCTION**

Dusty, complex, or colloidal plasma physics is part of the field of soft condensed matter, similar to granular material physics and the physics of colloidal suspensions[1,2]. In dusty plasma experiments, small particles, typically micrometer-sized, are immersed in plasma. Due to the collection of electrons and ions from the plasma, the particles obtain a large negative charge. As a result, they can be manipulated with electric fields and exhibit inter-particle interactions. Due to the large dust particle inertia, dusty plasma systems allow the study of solid state phenomena, fluid dynamics, turbulence and much more, on length- and timescales accessible with ordinary optical techniques[3,4].

In laboratory dusty plasma experiments, the strong electric field present in front of the powered electrode allows the dust particles to be levitated against the force of gravity. Theoretically, the charging of micro-particles in plasma is described by Orbital Motion Limited theory (OML)[5] and the electric field in which the particles are suspended is usually assumed to vary linearly with the height above the electrode[6]. Many experimental studies have been performed to determine the particle charge, most of which depend on perturbing the plasma parameters or perturbing the dust particles in the plasma, since probe measurements in dusty plasma systems are not reliable[7–9].

Two recent experiments, however, were performed to measure dust particle charge without perturbing the plasma parameters. The first employed a rotating electrode, causing rotation of the suspended particles due to rotational flow of the neutral gas[10]. Multiple particle sizes were used to probe different position within the discharge. This experiment concluded that the dust charge becomes *more negative than predicted by OML theory as the particles move closer to the electrode surface*. The second involved placement of the entire discharge cell onto a centrifuge, allowing the effective gravitational force to be enhanced and the particles to be pushed closer to the electrode surface[11]. This experiment concluded that the dust charge becomes *less negative as the particles move closer to the electrode surface*. Not only do the results from these experiments seem to contradict each other, they also bring the usual assumptions about dust particle charging and levitation in dusty plasma experiments into question. We will present a self-consistent fluid model[12] which shows that the two results above are in fact consistent with one another. This model allows us to compute the dust potential and, in turn, the dust charge as a function of height above a powered electrode by taking the effects of ion flow and electron depletion into account.

## II. DUST CHARGING INCLUDING ION FLOW AND ELECTRON DEPLETION

A dust particle in plasma collects ions and electrons, which, for typical laboratory settings, is the dominant charging process[13]. The particle will reach its equilibrium charge when the collected plasma-currents cancel:

$$I_e + \sum_i I_i = 0, \tag{1}$$

where the subscript $e$ denotes electrons and $i$ denotes the different ion species. In this paper, we consider ionized argon as the only ion species.

These currents can be calculated by integrating over the velocity distribution of the plasma species, $f_{e,i}(v_{e,i})$, including the cross section for capture of the plasma particle by the charged dust particle, $\sigma_{e,i}(v_{e,i})$, which is calculated using Orbital Motion Limited (OML) theory[5]:

$$I_{e,i} = \mp e \int v_{e,i} f_{e,i}(v_{e,i}) \sigma_{e,i}(v_{e,i}) d^3 v_{e,i}. \tag{2}$$

In the plasma above the electrode where the dust is levitated (usually assumed to be the sheath-region), the electron velocity is assumed to be distributed according to a Maxwellian distribution, characterized by the temperature $T_e$. The ions moving toward the electrode obtain a drift velocity $u_+$ in the local electric field, which is significant compared to the ion thermal velocity, $u_+ > \sqrt{k_B T_i / m_i}$. Their velocity distribution function can therefore be approximated by a *shifted Maxwellian distribution*[14],

$$f_e(v_e) = \left(\frac{m_e}{2\pi k_B T_e}\right)^{3/2} n_e \exp\left(\frac{-m_e v_e^2}{2 k_B T_e}\right), \tag{3}$$

$$f_+(v_i) = \left(\frac{m_i}{2\pi k_B T_i}\right)^{3/2} n_i \exp\left(\frac{-m_i (v_i - u_+)^2}{2 k_B T_e}\right). \tag{4}$$

Note that all plasma parameters, $n_e, T_e, n_i, T_i, u_+$ are functions of the height above the electrode, so $n_e = n_e(z)$ etc., but that we have now dropped the explicit notation. (We investigate the dust charge along the symmetry axis of the discharges in this paper, thus the radial dependence is ignored.) The density of Maxwellian electrons, for instance, is expected to drop off in the repulsive potential, $V(z)$, towards the lower electrode as $n_e(z) \propto \exp(eV(z)/k_B T_e(z))$, however, all profiles are self-consistently obtained using the fluid model as discussed below.

Plugging the distribution functions into equation (2), together with the OML cross sections, we find the electron and ion currents to a negatively charged dust particle to be

$$I_e = -e\sqrt{8\pi} a^2 v_{Te} n_e \exp(\Psi), \tag{5}$$

$$I_i = e\sqrt{2\pi} a^2 v_{Ti} n_i \left[\sqrt{\frac{\pi}{2}} \frac{1 + M_+^2 - 2\gamma\Psi}{M_+} \operatorname{erf}\left(\frac{M_+}{\sqrt{2}}\right) + \exp\left(\frac{-M_+^2}{2}\right)\right], \tag{6}$$

where $a$ is the dust particle radius, $\Psi = e\Phi_D/k_B T_e$ is the normalized dust particle surface potential ($\Psi < 0$), $v_{Tj} = \sqrt{k_B T_j/m_j}$ is the thermal velocity for species $j$, $M_+ = u_+/v_{Ti}$ is the ion thermal Mach number, and $\gamma = T_e/T_i$.

In the limit of low drift ($M_+ \ll 1$), the above equation reduces to the standard OML form for the plasma bulk,

$$I_i = e\sqrt{8\pi}a^2(1 - \gamma\Psi)n_i v_{Ti}, \tag{7}$$

while for very high drift ($M_+ \gg 1$), it approaches

$$I_i = e\pi a^2 \left(1 - \left[\frac{2k_B T_e}{m_i u_+^2}\right]\Psi\right)n_i u_+, \tag{8}$$

which corresponds to a much smaller ion current. The dust charge number, $Z$, i.e. the number of electrons residing on a given dust particle, can be found by assuming a capacitor model for the dust particle; $Z = 4\pi\epsilon_0 a k_B T_e \Psi/e^2$.

In this paper, we employ a self-consistent numerical fluid model to obtain the required profiles above the lower electrode and subsequently obtain $\Psi$ and $Z$ in this region. These results are then compared with results obtained in the literature. The next section briefly discusses the numerical model.

## III. FLUID MODEL AND CALCULATED PLASMA PROFILES

In order to calculate the dust charge for the two experiments mentioned above, we employed a self-consistent fluid model, which has been successfully used in the past for modeling microgravity dusty plasma experiments, as well as gravity-dominated experiments[12,15]. A brief description is given here.

The model solves the continuity equations for each plasma species including ionization source terms, $S_{e,i}$,

$$\frac{\partial n_{e,i}}{\partial t} + \nabla \cdot \mathbf{\Gamma}_{e,i} = S_{e,i}, \tag{9}$$

where the particle fluxes are found by employing a *drift-diffusion approximation* (rather than by solving the momentum equation)

$$\mathbf{\Gamma}_{e,i} = \mu_{e,i} n_{e,i} \mathbf{E} - D_{e,i} \nabla n_{e,i}, \tag{10}$$

where $\mu$ and $D$ are the mobility and diffusion coefficient, respectively.

For the electrons, the instantaneous electric field is used, which is found from Poisson's equation

$$\nabla^2 V = -\frac{e}{\epsilon_0}(n_e - n_i), \tag{11}$$

$$\mathbf{E} = -\nabla V. \tag{12}$$

The high ion inertia is taken into account using an effective electric field in the ion flux equation, found by iterating

$$\frac{dE_{Eff}}{dt} = v_i(E - E_{eff}), \tag{13}$$

with $v_i = e/m_i\mu_i$.

The ion energy is assumed to be locally dissipated through frequent collisions with the neutrals. The electron energy is solved using a similar drift diffusion scheme for the particle flux,

$$\frac{\partial w_e}{\partial t} + \nabla \cdot \Gamma_{w_e} = J_e \cdot E + S_{w_e}, \tag{14}$$

where $J_e \cdot E$ is the Ohmic heating term and $S_{w_e}$ are sink terms for inelastic collisions, including electron-impact excitation and ionization of argon. The heat flux is given by

$$\Gamma_{w_e} = -\frac{5}{3}\mu_e w_e E - \frac{5}{3}D_e \nabla w_e. \tag{15}$$

In these equations $w_e = n_e \epsilon$ is the electron energy density, with $\epsilon$ the average electron energy.

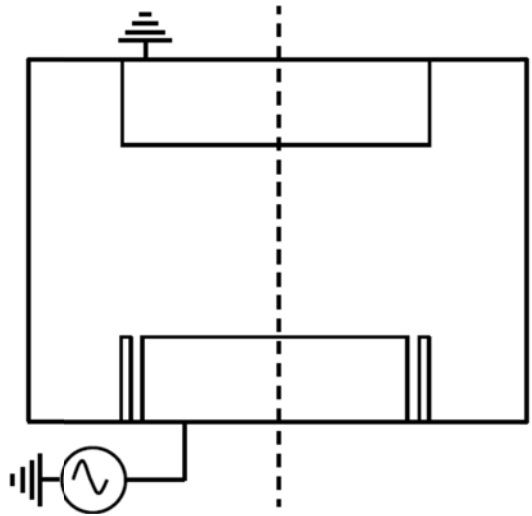

FIG. 1. A sketch of the discharge geometry considered in the model. Results are shown along the symmetry axis above the lower electrode, for half the inter-electrode gap. A ground shield is also included.

These equations are solved on a grid and iterated in time using sub-RF time-steps until convergence is reached, in which case the solutions become periodic over one RF-cycle. Figure 1 shows the general geometry of the discharge chambers modeled. The model assumes cylindrical symmetry.

For the experiments modeled here, we consider only the plasma above the lower electrode (all experiments considered are asymmetrically powered on the lower electrode) and show results along the symmetry axis, for half the inter-electrode gap distance. To model different experiments, the dimensions of the discharge chamber are varied, including the electrode size, inter-electrode distance, and outer wall dimensions.

In order to radially confine the dust, depressions in the electrode or rings on top of the electrode are used, resulting in parabolic potential energy wells for the dust. Rather than complicating the geometry of the lower electrode in the model, an additional confinement potential is added on the lower electrode as a boundary condition, which falls off quadratically with the distance from the symmetry axis over the width of the simulated electrode cutout.

The relevant profiles were calculated for the geometry of the rotating electrode method experiment. The pressure is fixed at 20 Pa, but the driving potential amplitudes, $V_{RF}$, were varied. In the actual experiment, the pressure was set at 4 Pa, but the fluid model is not applicable at such low pressure settings. Nonetheless, the results obtained are consistent with the observations made in the experiment as shown below. Furthermore, the effects of electron depletion and ion flow, which are included in the fluid model, continue to be important at low pressures.

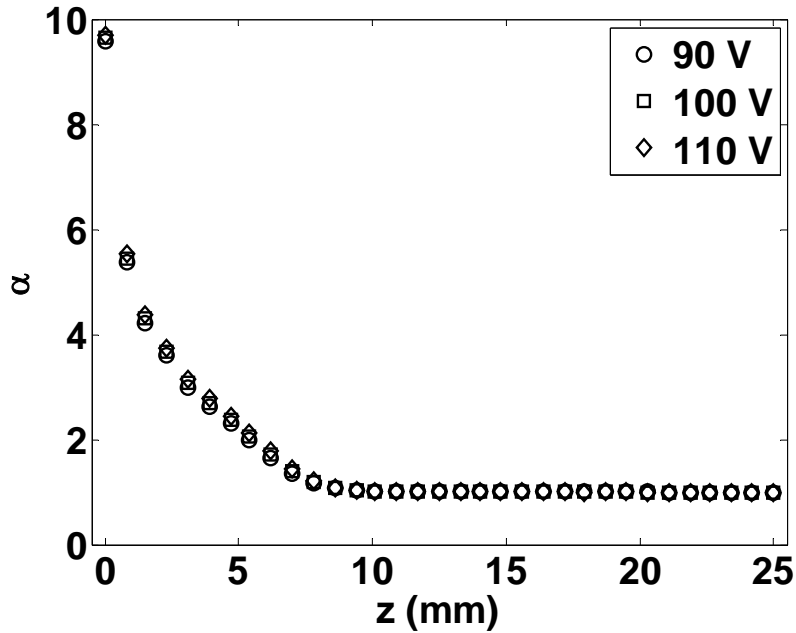

FIG. 2. The ratio of the ion density to the electron density above the lower electrode for various driving potentials, $V_{RF}$, at 20 Pa.

Figure 2 shows the density ratio $\alpha(z) = n_i(z)/n_e(z)$. The electron density quickly becomes an order of magnitude smaller than the ion density in the plasma above the electrode, independent of driving potential amplitude, due to the repulsive potential of the wall with respect to the plasma.

The ion Mach number, $M_+$, is presented in figure 3. The ion drift velocity is obtained from equation (10) as $\boldsymbol{u}_+(z) = \boldsymbol{\Gamma}_i(z)/n_i(z)$. The ion flow speed increases towards the lower electrode and increases roughly linearly with $V_{RF}$.

The $T_e(z)$ profile varies as well, as shown in figure 4. In the fluid model, the electron temperature is defined through the average electron energy, $\epsilon(z) = w_e(z)/n_e(z) = 3k_B T_e(z)/2$. The local variation in electron temperature is therefore a reflection of both local density variations, as well as inelastic collision processes leading to energy loss.

Once these profiles are obtained, the charging current equations can be solved and from the current balance equation the dust surface potential profile, $\Phi_D(z)$, can be calculated. The profile is shown in figure 5 for the geometry of the rotating electrode method experiment. Clearly, $\Phi_D(z)$ varies strongly throughout the plasma above the electrode and has a clear minimum for the geometry considered here. The dust potential attains a value of -7.6 V

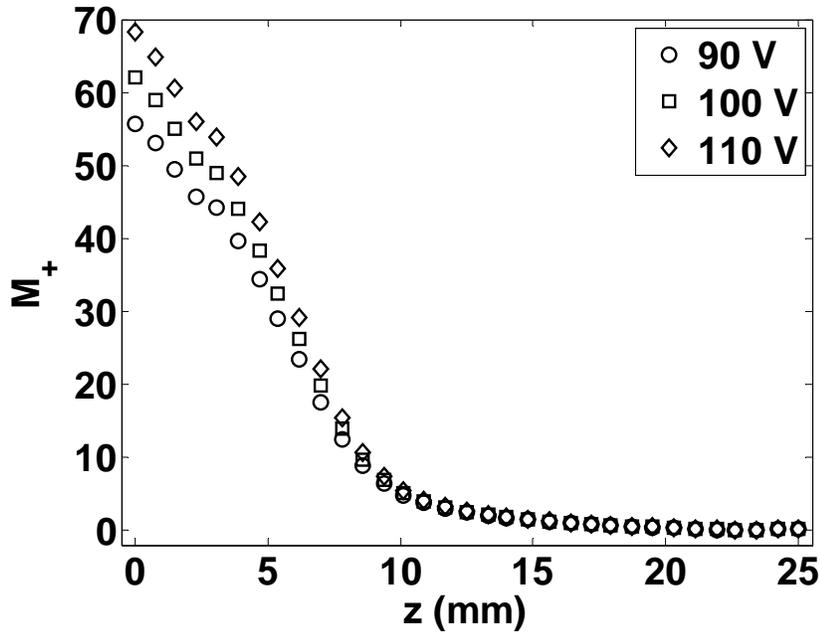

FIG. 3. The ion Mach number profile above the lower electrode for various driving potentials, $V_{RF}$, at 20 Pa.

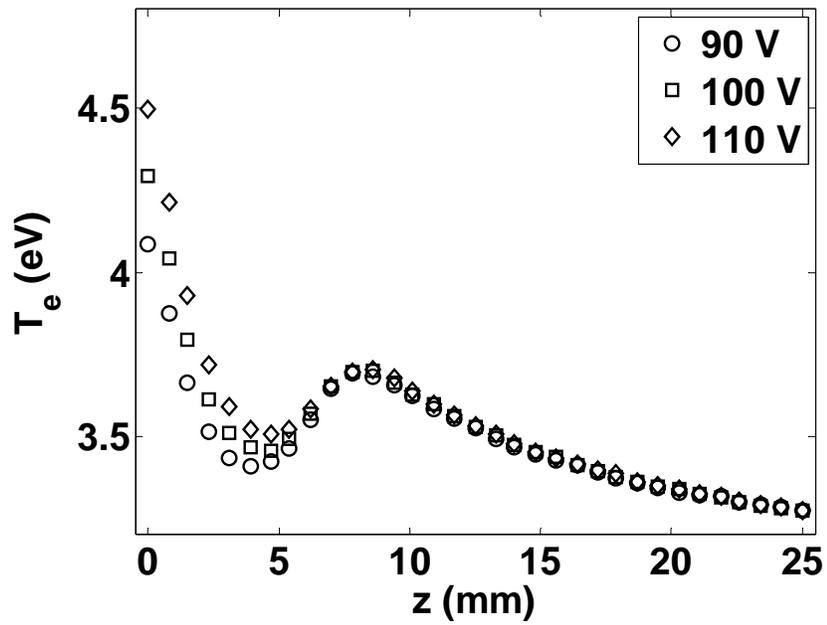

FIG. 4. The electron temperature profile in the plasma above the lower electrode for various driving potentials, $V_{RF}$, at 20 Pa.

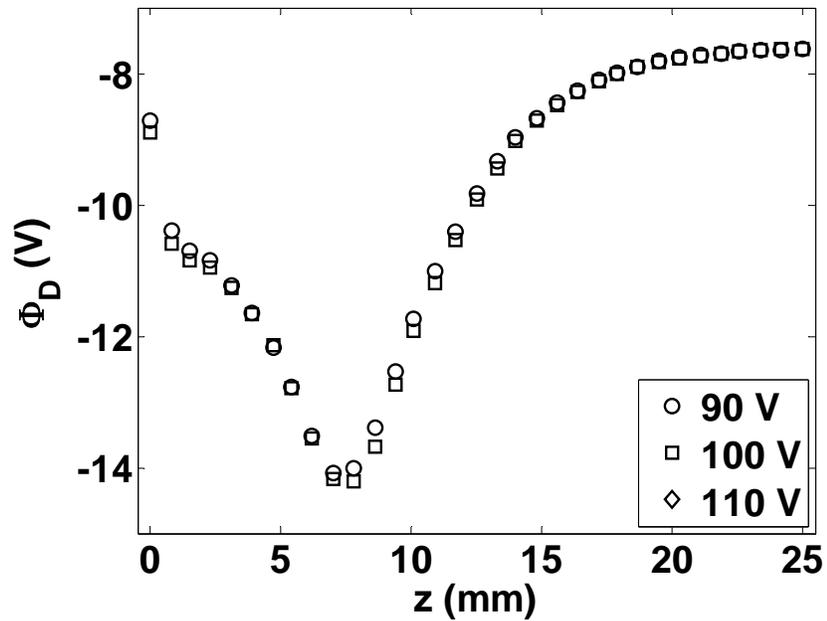

FIG. 5. The dust potential, $\Phi_D(z)$, as a function of the height above the lower electrode as determined from the fluid model. Profiles are shown for various RF voltages at 20 Pa.

Dust particle charge in plasma with ion flow and electron depletion in the bulk and -8.9 V near the lower electrode. These values are equivalent to the dust potential found using equations (7) and (8). Note that $\Phi_D(z)$ is independent of particle size, but that the dust charge number, $Z$, depends on the dust particle size through the capacitor model.

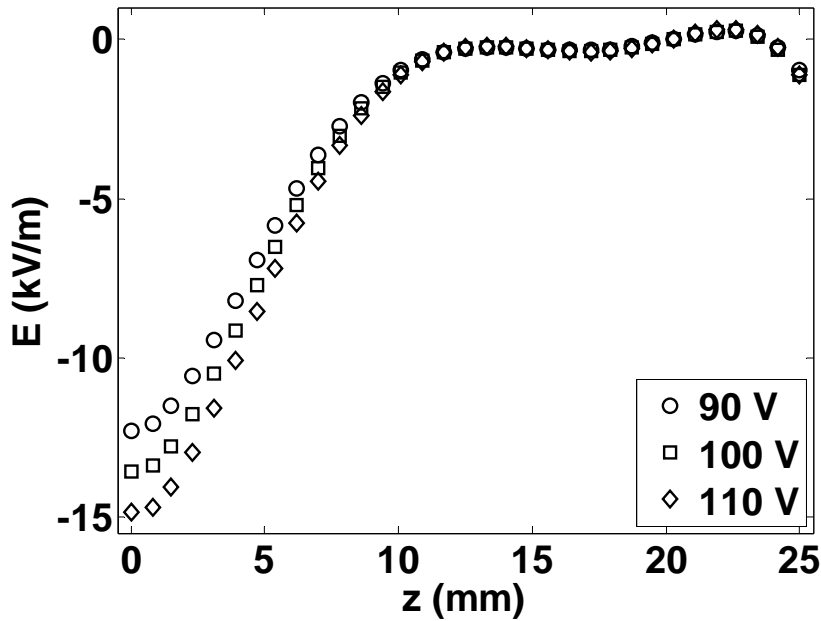

FIG. 6. The electric field profile for various RF voltages at 20 Pa as determined from the fluid model.

Combining the dust charge with the model solution for the electric field, $\boldsymbol{E}(z)$, shown in figure 6, the electric force is then found as $F_E(z) = Q_D(z)E(z)$. Assuming balance with gravity, $F_g = m_D g$, the levitation height of the dust is found. The next section compares the obtained dust charge and levitation heights with the results reported in literature. Note that at the position where the electric field starts to increase strongly (at $z \approx 10\ mm$), $\Phi_D(z)$ also becomes more negative. This shows the effect of the increasing ion drift in the growing electric field on the dust charge. Then, when the electron depletion becomes significant (at $z \approx 5\ mm$), the dust charge becomes less negative again, showing the effect of the decreased electron density on the electron contribution to the charging currents. This behavior is qualitatively consistent with a similar figure reported in the literature[16].

**IV. COMPARISON TO THE EXPERIMENTS**

In this section we will discuss the conclusions regarding the dust charge number profile found in the aforementioned experiments. Our fluid model will be applied to each experiment, and the dust charge number dependence on height above the lower electrode will be obtained.

99

## A. Rotating Electrode Method

At a pressure of 4 Pa and RF voltage of 100 V, Carstensen et al.[10] found that 12 μm and 20 μm diameter particles had a charge number, $Z$, of 19500 and 60300, respectively. The increase in charge found on the 20 μm particle is larger than predicted from the capacitance model at constant $\Phi_D(z)$, which states that the particle charge should increase linearly with the particle radius. Therefore, it was determined that the dust potential, $\Phi_D(z)$, was not constant in the sheath, but rather became more negative with decreasing height above the lower electrode, causing the dust charge to also become more negative.

The results obtained with our fluid model at a pressure of 20 Pa are shown in Table I where a nonlinear relationship between $Z$ and particle radius can be observed. In addition, we find that $Z$ varies with RF voltage and hence with increased power applied to the discharge. Table II shows the charge number ratio for different particle sizes at different RF voltages. The theoretical value calculated using the ratio of the particle radii is also shown. We find the charge number ratio to be larger than or equal to the theoretical value, which is consistent with the reported results.

TABLE I. The charge number, $Z$, for various particle sizes and RF voltages at 20 Pa.

| Diameter (μm) | $Z$ (90 V) | $Z$ (100 V) | $Z$ (110 V) |
| --- | --- | --- | --- |
| 12 | 53539 | 53542 | 53574 |
| 16 | 77278 | 77702 | 77794 |
| 20 | 96762 | 98711 | 99750 |

TABLE II. The charge number ratio for different particle sizes and RF voltages at 20 Pa. Theoretical values are the ratio of the particle radii.

| $V_{RF}$ (V) | $Z_{16}/Z_{12}$ | $Z_{20}/Z_{16}$ | $Z_{20}/Z_{12}$ |
| --- | --- | --- | --- |
| 90 | 1.44 | 1.25 | 1.81 |
| 100 | 1.45 | 1.27 | 1.84 |
| 110 | 1.45 | 1.28 | 1.86 |
| Theoretical value | 1.33 | 1.25 | 1.67 |

Dust particle charge in plasma with ion flow and electron depletion

**B. Hypergravity Experiment**

Beckers *et al.*[11] recently performed an experiment in which a centrifuge was used to subject microparticles to hypergravity conditions. As expected, the resulting increase in the apparent gravity, indicated by the acceleration $g^*$, caused the particles to levitate closer to the lower electrode. Using this method, they were able to nonintrusively measure the electric field strength and the particle charge above the powered electrode. In the experiment it was determined that the dust charge number decreased as the particles levitated closer to the lower electrode.

The geometry of the cell along with the discharge parameters used in the experiment were inserted into the fluid model and the relevant plasma parameters were calculated. Figure 7 shows the equilibrium height for a 10.2 µm diameter particle as a function of the apparent gravitational acceleration, $g^*$. The equilibrium height was determined from the force balance between the apparent gravitational force and the electric force. Consistent with the experiment, the model shows that the particle levitates closer to the lower electrode as $g^*$ is increased.

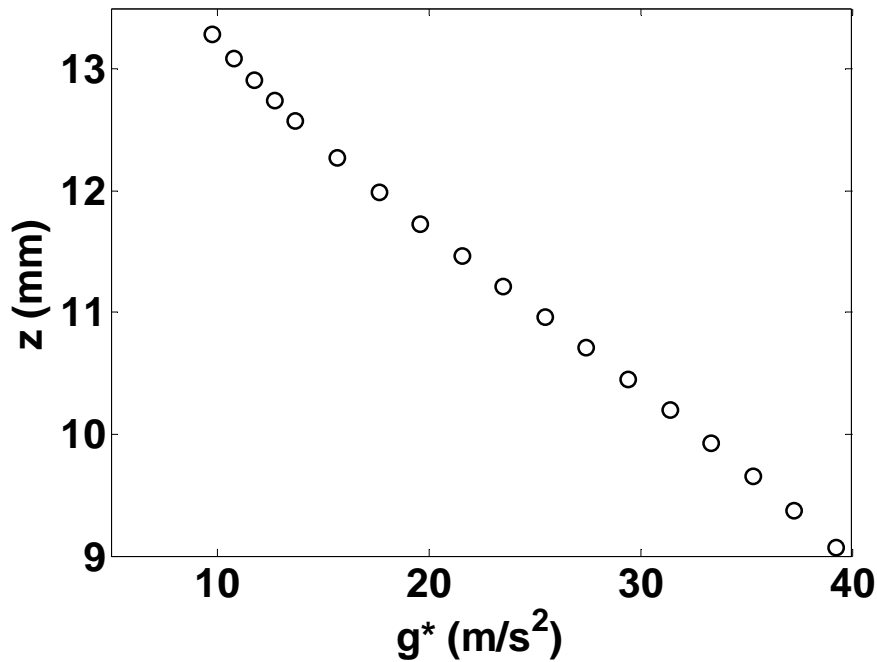

FIG. 7. The levitation height of a 10.2 µm particle as a function of applied apparent gravitational acceleration, $g^*$ at 20 Pa and 80 V.

The dust charge number and the electric field strength as determined from the fluid model are shown in figure 8 with symbols marking the levitation heights at various values of $g^*$. For the 14 and 16 μm diameter particles, no force balance was obtained for $g^* = 3g$.

The dust charge number reaches a maximum value, $Z_{max}$, at a particular position, $z(Z_{max})$. The fluid model indicates that the 6 μm particles always levitate above $z(Z_{max})$ for all values of $g^*$ shown and when the apparent gravitational acceleration is increased from $0.5g$ to $3g$, the charge number also increases. But for larger particles, which levitate closer to the lower electrode, an increase in the apparent gravity causes the charge on the particles to increase for small values of $g^*$ and decrease for larger values of $g^*$. All particles will therefore levitate closer to the lower electrode as the apparent gravity is increased, but the particle charge can increase or decrease, depending on their position relative to $z(Z_{max})$.

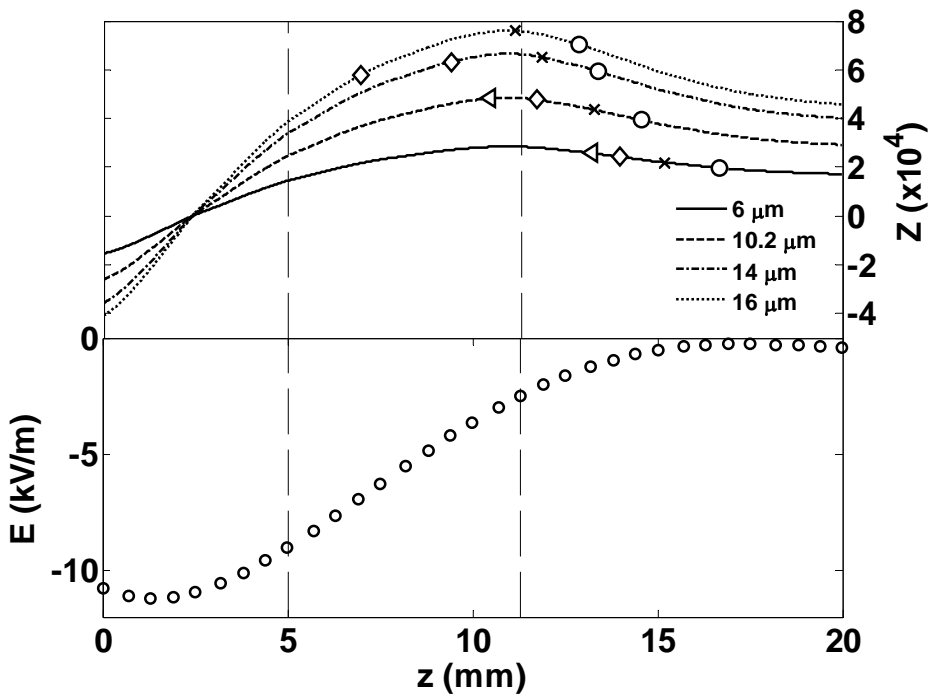

FIG. 8. $Z$ as a function of height above the lower electrode for four particle diameters. The symbols o, x, ◊ and represent levitation heights at $g^* = 0.5g$, $1g$, $2g$, and $3g$, respectively. The bottom graph shows the electric field profile obtained with the fluid model.

Beckers et al. found the electric field to be linear in the region of the discharge they probed. The electric field obtained with the fluid model contains a linear region as well, as indicated by the two vertical dashed lines in figure 8. It can be seen that particles levitated within this region would indeed show a decrease in the dust charge as $g^*$ is increased and hence $z$ is decreased.

Figure 8 also shows that the charge number for all particle sizes is equal to zero at $z = 2.4\ mm$ and then becomes positive close to the lower electrode. While the change in sign of the dust charge number should occur for $z < 2.4\ mm$, the shape of the profile will be different than that shown. The current equations (5) and (6) should be solved assuming a positive dust charge to accurately model the dust charge number in this region.

## V. DISCUSSION AND CONCLUSIONS

The dust surface potential as a function of height in the plasma above a powered electrode has been determined by taking into account electron depletion and ion flow in the charging equations for a dust particle. Using a plasma fluid model to obtain the relevant plasma parameter profiles, the charging equations were solved allowing calculation of the dust surface potential. This potential is not constant throughout the plasma above the electrode, but instead decreases strongly with decreasing height, due to the increased downward ion flow near the sheath-bulk boundary. Even closer to the electrode, electron depletion becomes significant and the dust surface potential becomes less negative. The combined result is a local minimum of the dust surface potential and hence a local maximum of the dust charge number, $Z_{max}$, at a height $z(Z_{max})$. Therefore, the standard approach which uses bulk plasma values in OML theory to describe the dust charge in laboratory dusty plasma experiments is clearly not valid.

These results explain the apparent contradiction implied by the two recent experimental papers discussed above. From our simulations, we conclude that the rotating electrode method experiment probed locations where $z > z(Z_{max})$, and therefore the charge number increased as the particle height decreased. On the other hand, the hyper-gravity experiment probed locations where $z < z(Z_{max})$, so the charge number decreased as the height decreased. Hence, there is no actual contradiction between their results since each experiment probed different regions of the dust charge number profile in their discharge, due to the differences in their operating parameters, and the geometry of their experimental setup.

Even though the charging equations used in this paper are simple, the qualitative results are in accordance with the reported experiments. Any additional quantitative differences between the results reported here and the results reported in the literature are likely due to approximations and limitations of the fluid model. Effects not present in our model include non-linear electron heating and secondary electron emission from the electrode which can be important for the ionization in the discharge at low pressures[17,18]. This limits our ability to model discharges at lower pressures. Furthermore, the inclusion of these effects would result in additional energetic electrons, which could result in a more negative dust charge than presented here. At higher pressures, the inclusion of ion collisionality[19] and trapped ions[20] would increase the ion flow to dust particles, which could result in a more positive dust charge. Even though these effects may be important, we expect them to be secondary to the effects of electron depletion and ion flow, especially at the pressures considered in this paper.

Finally, it is important to note that even though we applied the effect of electron depletion and ion flow to the charging of particles above a powered electrode, other regions where these effects occur will show similar dust charge variation. For instance, systems containing regions of ambipolar electric fields in front of floating discharge boundaries, such as the glass tube in the PK-4 experiment currently being investigated[21], should also show charge variation. This should also be seen in the void boundary in microgravity dusty plasma experiments[22], since this region definitely shows electron enhancement and depletion, as well as ion flow effects.

Knowledge of the electric field and dust charge profiles in systems where electron depletion and ion flow are significant is essential to understanding dust particle behavior. Two of such systems we intend to study by including the effects discussed in this paper are vertical chains of dust particles and complex plasma bilayers. Of particular interest in these systems is the interaction between the particles which strongly depends on their charge as a function of height above the electrode.

## ACKNOWLEDGMENTS

This material is based upon work supported by the National Science Foundation under
Grant No. 0847127 and support by the Texas Space Grant Consortium through a graduate fellowship.